\documentclass[a4paper,12pt]{article}
\usepackage{epsf,amsmath}
\usepackage{indentfirst}
\usepackage{amssymb}
\usepackage{hhline}
\usepackage{textcomp}
\usepackage{array,tabularx}
\def\@biblabel#1{#1.}
\makeatother
\begin{document}

\title{\Large\bf Type Ia Supernovae: Non-standard Candles of the Universe}

\author{A. I. Bogomazov$^1$, A. V. Tutukov$^2$ \\
\small\it $^1$Sternberg Astronomical Institute, Moscow State
University, \\ \small\it Universitetskii pr. 13, Moscow, 119991 Russia \\
\small\it $^2$Institute of Astronomy, Russian Academy of Sciences,
\\ \small\it Pyatnitskaya ul. 48, Moscow 119017, Russia}

\date{\begin{minipage}{15.5cm} \small
We analyze the influence of the evolution of light absorption by
gray dust in the host galaxies of type Ia supernovae (SNe Ia) and
the evolution of the mean combined mass of close-binary
carbon-oxygen white dwarfs merging due to gravitational waves (SNe
Ia precursors) on the interpretation of Hubble diagrams for SNe
Ia. A significant increase in the mean SNe Ia energy due to the
higher combined masses of merging dwarfs should be observable at
redshifts $z > 2$. The observed relation between the distance
moduli and redshifts of SNe Ia can be interpreted not only as
evidence for accelerated expansion of the Universe, but also as
indicating time variations of the gray-dust absorption of light
from these supernovae in various types of host galaxies,
observational selection effects, and the decrease in mean combined
masses of merging degenerate dwarfs.
\end{minipage} } \maketitle \rm

Astronomy Reports, v. 55, pp. 497-504 (2011)

\section{INTRODUCTION}

This paper continues our study \cite{bogomazov2009}, where we
demonstrated that the mean combined masses of binary carbon-oxygen
white dwarfs (CO WDs) merging due to gravitational-wave radiation
should evolve with time. Paper \cite{bogomazov2009} presents the
results of population syntheses performed with the ``Scenario
Machine'' (a computer code for evolutionary studies of close
binaries), as well as computations of the merger rate and mass
distributions of merging white dwarfs (WDs). Such studies are
important because observations of distant type Ia supernovae (SNe
Ia) in the late 1990's led to the conclusion that the expansion of
the Universe was accelerating\footnote{At $z\approx 1$ were found
to be fainter than expected for standard supernovae of this type
by, on average, about $0^{m}.2$
\cite{riess1998a,perlmutter1999a}.}
\cite{riess1998a,perlmutter1999a}. Thus, understanding the physics
of SNe Ia determining the possible evolution of their brightness
with time is needed to understand the fundamental properties of
the Universe. In addition, in contrast to our first study, we also
consider here the influence of the evolution of gray dust in the
host galaxies of SNe Ia on changes in the absorption of light from
these supernovae. Light absorption by gray dust is able to produce
the same effect of fainter distant supernovae as the proposed
accelerated expansion of the Universe.

There are two main models for the origin of a SNe Ia outburst. The
first requires a close binary consisting of a CO WD and a
non-degenerate star, whose matter is accumulated onto the white
dwarf via accretion and hydrogen and helium burning in shell
sources in the dwarf's envelope. As soon as the WD mass exceeds
the Chandrasekhar limit, a thermonuclear explosion occurs (or an
explosion of the helium shell causes detonation of carbon in a
dwarf with a mass below the Chandrasekhar limit), observable as a
SN Ia \cite{whelan1973}. The second model requires a close binary
consisting of two CO dwarfs. Under the influence of
gravitational-wave radiation, the binary's semi-major axis
decreases, the dwarfs merge, and the resulting thermonuclear
explosion is manifest as a SN Ia
\cite{iben1984,tutukov1980}\footnote{In this case, the combined
mass of the dwarfs probably exceeds the Chandrasekhar limit.}. We
consider the second scenario to be the main one
\cite{yungelson2010}.

Yungel'son \cite{yungelson2010} aims to prove that mergers are the
principal, and possibly the only, source of SNe Ia. The accretion
scenario is probably unable to explain SNe Ia: CO WDs are probably
not able to accumulate enough gas to make their masses exceed the
Chandrasekhar limit \cite{yungelson2010}. Observations also
present evidence in favor of the WD-merger scenario
\cite{stefano2010,gilfanov2010}. However, the viewpoint of
\cite{stefano2010,gilfanov2010} are not shared by all authors
\cite{lipunov2010}. Nevertheless, the main mechanism for SN Ia
explosions in elliptical galaxies with ages in excess of
approximately 200 million years is CO WD mergers
\cite{gilfanov2010}-\cite{jorgensen1997}. The energies of some SN
Ia correspond to masses of burned carbon and oxygen $\approx 2
M_{\odot}$ \cite{silverman2010}, providing direct evidence in
favor of the dwarf-merger model and confirming our estimates
\cite{bogomazov2009}.

The merger rate of close binary WDs and its dependence on the
star-formation history were studied in
\cite{tutukov1994,jorgensen1997}; the results of later studies are
not very different (see, for example,
\cite{ruiter2009,mennekens2010}). This may be because the WD
merger rate is virtually independent of the merger mechanism if
the merger time scale depends only on the initial semi-major-axis
distribution for binary WDs \cite{lipunov2010}. Merger candidates
are actively being searched for, and merger rates and the delay of
the merger of degenerate close-binary components estimated (see,
for instance, \cite{kilic2010,maoz2010}).

The role of the possible evolution of the energy of SNe Ia was
studied ten years ago in \cite{drell2000}, where several possible
hypothetical luminosity evolutions for SNe Ia were considered.
According to those computations, introducing even a small amount
of evolution makes it very difficult to determine cosmological
parameters such as the dark-energy density. We demonstrated in
\cite{bogomazov2009} that SNe Ia can be considered standard
candles only after approximately one billion years of galaxy
evolution\footnote{This result is independent of the
star-formation history for the three versions of the star-forming
function considered in \cite{bogomazov2009}, all stars are born
simultaneously (star formation in an idealized elliptical galaxy),
a constant star-formation rate (star formation in an idealized
spiral galaxy), and the star formation history of
\cite{kurbatov2005}.}. In the course of their evolution, the mean
energy of SNe Ia should decrease by approximately 10\%, and the
difference between the highest and lowest energies of SNe Ia
should be at least a factor of 1.5\footnote{It is a factor of two
in the extreme case: a merger of two WDs with a combined mass
approximately at the Chandrasekhar limit, or a merger of two WDs,
each having a mass close to the Chandrasekhar limit. However,
mergers of CO WDs with combined masses in excess of $2.1
M_{\odot}$ are rare. In addition, if we take into account the
hypothetical possibility of explosions from CO WD mergers with
combined masses below the Chandrasekhar limit, the difference
between the highest and lowest energies of SNe Ia could be even
greater than a factor of two. Studies of possible
sub-Chandrasekhar scenarios for the origin of SNe Ia have now
begun (see, for instance, \cite{sim2010,ruiter2010}).  } at both
high and low redshifts. These obvious observational selection
effects must be taken into account when estimating the parameters
of the Universe's expansion using SNe Ia. The maximum brightness
of such supernovae is not a universal constant, and has some
dispersion. While we see ``all'' the closest SNe Ia, only the
brightest are detected at increasing distances. This could imitate
a brightening of SNe Ia with distance. The magnitude of this
effect is comparable to the intrinsic brightness dispersion for
this type of supernovae.

The brightness of a SNe Ia can be attenuated by absorption in
matter around the supernova precursor, as well as absorption by
dust in the supernova's parent galaxy; the principal
characteristics of dust in the host galaxies of supernovae, total
dust mass, distribution of dust in the galaxy, and dust
composition can all vary in time. Redshift-dependent absorption by
intragalactic gray dust could introduce errors into the extent to
which SNe Ia can be considered standard candles; not taking into
account absorption by gray dust makes photometric distances too
large compared to a dust-free Universe and could give rise to an
apparent ``acceleration'' of the expansion of the Universe
\cite{holwerda2008}. According to \cite{corasaniti2006}, taking
into account absorption by gray dust can make a SNe Ia fainter by
$0.^{m}08$ at a redshift of $z=1.7$, potentially capable of
causing disagreement between Hubble diagrams calculated assuming a
dark-energy density $\Omega_{\lambda}\simeq 0.7$ and observations
of SNe Ia.

Absorption by intergalactic gray dust was suggested in
\cite{aguirre1999a,aquirre1999b} as an explanation of the fainter
magnitudes of supernovae at large distances, as an alternative to
accelerating expansion of the Universe, almost immediately after
the classic papers \cite{riess1998a,perlmutter1999a} on
observations of distant supernovae. In addition, a model with
intergalactic dust was considered in \cite{goobar2002}, where it
was suggested that intragalactic gas was not gray, and so could be
accounted for by means of precision spectroscopy. However, later
studies gave somewhat different results \cite{gorbikov2010}.

A comparison of the brightnesses of some 85 thousand quasars and
the positions of about 20 million galaxies (based on the SDSS
survey) demonstrated the presence of dust in intergalactic space,
comparable in amount and composition to the dust in galactic disks
\cite{menard2010a}. Problems related to the correction of
supernova parameters taking into account absorption by this dust
are discussed in \cite{menard2010b}, where the influence of gray
dust is not considered. The authors of
\cite{menard2010a,menard2010b} note that if the detected dust does
not contain gray dust, they are able to provide an upper limit for
absorption, but if gray dust is present, their computations give a
lower limit for the influence of dust.

The example of the Milky Way shows that absorption in the parent
galaxy can reach $1^{m}$, even in the polar direction. It also
makes supernovae in disk galaxies fainter. For example,
observations of stars in the Sloan Digital Sky Survey (SDSS) led
to the discover and investigation of three absorbing clouds in the
Galaxy, each hundreds of parsecs in size. The absorption in these
clouds is $0^{m}.2-0^{m}.4$, and the clouds exhibit gray, almost
wavelength-independent absorption in the spectral range of the
survey \cite{gorbikov2010}. When estimating the evolution of the
brightness of SNe Ia in time, we must remember that the gas
component of the galaxies containing these supernovae also
evolves. This introduces additional uncertainty into brightness
estimates for supernovae. The higher infrared luminosities of
distant ($z\approx 1$) galaxies indicate the presence of high dust
contents in young galaxies (e. g., \cite{buat2010}).

We can put the following limit on radiation absorption by dust in
an elliptical galaxy with a mass comparable to that of the Galaxy.
Let the total mass lost by red supergiants be $\sim 1 M_{\odot}$
per year. The mass loss time scale for an elliptical galaxy is
$\sim 10^8$ years. Thus, the stationary amount of gas in the
galaxy is $\sim 10^8 M_{\odot}$. In this case, the absorption for
the usual abundance of dust will be $\approx 0^m.1$, if we assume
the same efficiency for dust absorption in elliptical and disk
galaxies.

To estimate the absorption of light from supernovae in disk
galaxies, we applied the evolutionary model of
\cite{firmani1992}-\cite{kabanov2011}. The assumed mass and radius
of the model galaxy were $2\cdot 10^{11} M_{\odot}$ and $\sim
10^4$ pc. We estimate the star-formation rate from the condition
that the gaseous disk be fully ionized by young massive stars
\cite{firmani1992}. The thickness of the gas disk is estimated
from the condition that the rate at which energy is supplied to
the turbulent motion of the galaxy's gas by supernovae (II and
Ib,c) be the same as the dissipation rate of the turbulent motion
of this gas \cite{firmani1992}. In this case, the rate of
supernova outbursts is determined by the star-formation rate in
the model galaxy, and the dissipation of the turbulent motion of
the interstellar gas is determined by collisions between
interstellar gas clouds.

Modeling demonstrated that the star-formation rate in a disk
galaxy varies considerably in time \cite{wiebe1998}. The
star-formation rate increases during the several first billion
years. In a galaxy similar to ours, the star-formation rate
reaches a maximum of $\sim 10^2 M_{\odot}$ per year at an age of
$2- 4$ billion years. During the subsequent evolution, the
star-formation rate gradually decreases to its current level. The
model takes into account gas enrichment with heavy elements,
making it possible to follow the time evolution of the gas
component's optical depth when checking the mass of the gaseous
component. Knowing this, we are able to estimate the time
evolution of the apparent brightnesses of supernovae, assuming
that all are located in the planes of their host galaxies, which
are orthogonal to the line of sight. In this way, we can take into
account the effect of absorption of light from supernovae in disk
galaxies. For the current epoch, this amounts to $\approx 0^m.5$,
and increases for young galaxies with high gas and dust contents.
Dust is thus able to attenuate the brightness of SNe Ia.

It is especially important to note that we are concerned here with
gray dust, whose absorption is wavelength-independent. The
existence of such dust in the galactic plane is now beyond doubt
\cite{gorbikov2010}. Fine dust causes reddening of light and thus,
in principle, can be taken into account. It should also be borne
in mind that modeling of the time evolution of a galaxy's optical
depth remains uncertain, and the optical depth itself cannot be
estimated better than to within a factor of two
\cite{firmani1992,wiebe1998}. Gray-dust light absorption occurs
independently of the supernova model that is preferred. We will
assume here that the amount of gray dust in a parent galaxy of a
SN Ia is proportional to its total amount of dust.

\section{MODEL EVOLUTION OF SNe Ia
BRIGHTNESS WITH INCREASING REDSHIFT}

In this study, we analyze the computational results used to plot
Figs. 1-3 in \cite{bogomazov2009}\footnote{The computations of
\cite{bogomazov2009} were based on the ``Scenario Machine''. A
description of the code can be found in
\cite{lipunov1996,lipunov2009}, and of the population synthesis,
in \cite{popov2007}.}. We used these computations to derive the
mass distributions of merging WDs \cite{bogomazov2009} (Fig. 1)
and the dependence of the mean mass of merging WDs on the time
elapsed since the onset of star formation \cite{bogomazov2009} (
Fig. 2). Virtually independent of the form of the star-formation
functions, the mean mass of merging CO WDs with combined masses in
excess of the Chandrasekhar limit decreases from $\approx 1.9
M_{\odot}$ to $\approx 1.7 M_{\odot}$ within about one billion
years after the onset of star formation, after which the mean mass
of merging WDs remains practically the same.

The aim of this study is to provide a theoretical analog of a
Hubble diagram \cite{riess1998a} (Figs. 4, 5) taking into account
the brightness evolution of SN Ia and the evolution of dust
absorption, using the theoretical model presented in
\cite{kabanov2011}. We assume that the absolute magnitude of a SN
Ia depends on the mean mass of the merging CO WDs as

\begin{equation}
M=C - 2.5\cdot \log M_{\Sigma}, \label{magabs}
\end{equation}

\noindent where $M$ is the absolute magnitude of the SN Ia,
$M_{\Sigma}$ the combined mass of the merging CO WDs, and $C$ is a
constant. Thus, the higher the combined mass of the merging
dwarfs, the brighter the supernova. Using \ref{magabs} to
re-calculate the data plotted in Fig. 2 of \cite{bogomazov2009},
we obtained the dependence of the absolute magnitude of a SN Ia on
the time since the onset of the formation of the corresponding
stellar population.

The resulting dependence of the absolute magnitude on the time
from the onset of the galaxy's evolution until the supernova
outburst must be transformed into the dependence of the observed
SNe Ia magnitudes on the observed redshifts for the host galaxies.
These computations used Eq. 20 (to calculate the time-redshift
relation) and 28 (to calculate the photometric distance-redshift
relation) from \cite{sahni2000}. The free parameters in these
formulas are the density of dark energy $\Omega_{\lambda}$, the
matter density $\Omega_{M}$, the total density of the Universe
$\Omega_{tot}$, and the current Hubble constant $H_{0}$. Another
free parameter of our problem is the redshift $z^{*}$
corresponding to the onset of star formation in galaxies.

The last stage of our computations is to calculate the dependence
of $\Delta (m-M)$ on redshift $z$:

\begin{equation}
\Delta (m-M)=(m_1-C)-(m_2-C)=m_1-m_2, \label{deltam}
\end{equation}

\noindent where $m$ and $M$ are the apparent and absolute
magnitudes, $m_1$ the apparent magnitude calculated taking into
account the time evolution of the mean mass of the merging WDs,
and $m_2$ the apparent magnitude calculalated assuming that a SN
Ia is a standard candle with absolute magnitude $C$.

We took the dependence for light absorption by gray dust from
\cite{kabanov2011} (Fig. 2 presents the dependence of $A_v$ on the
redshift $z$; we used the curve corresponding to a disk galaxy
with the mass-radius relation $M\sim R^2$, see \cite{kabanov2011}
for details). The redshift dependence of the absorption in this
figure was derived using the formula $t(z)=\frac{2}{3 H_0
(1+z)^{3/2}}$ relating the time $t(z)$ elapsed since the Big Bang
and the redshift $z$. We place this relation on the Hubble diagram
in order to qualitatively illustrate the evolution of dust
absorption in disk galaxies, compared to the accelerated fading of
supernovae at redshifts $0.5-1$. Note also that absorption by dust
grains of any size was considered in \cite{kabanov2011}; that
paper contains no estimate of the fraction of gray absorption due
to grains with sizes in excess of $1\mu m$.

\section{RESULTS}

The results of our computations are presented in Figs. \ref{ris1}
and \ref{ris2}. The values computed using \ref{deltam} are plotted
versus redshift. The magnitude $m_2$ used in \ref{deltam} to
calculate the curves in Figs. \ref{ris1} and \ref{ris2} is the
dependence of the apparent magnitude $m$ on the redshift $z$,
assuming that all SN Ia are standard candles with absolute
magnitude $C$ and that they do not undergo any absorption (at
least no time-variable absorption). This relation corresponds to
the horizontal straight line at zero magnitude (curve 3). The
magnitude $m_1$ is the dependence of $m$ on $z$ taking into
account one of the studied effects.

Curve 1 describes the dependence of $\Delta (m-M)$ on $z$,
assuming that either absorption plays no role or it is constant in
time, and that the absolute magnitude evolves in time from its
maximum value at the beginning of the evolution to its minimum,
approximately constant, value after about a billion years of
stellar evolution in galaxies. The adopted cosmological parameters
are $\Omega_{\lambda}=0$, $\Omega_{M}=0.2$, $\Omega_{tot}=0.2$,
$H_{0}=70$ km s$^{-1}$ mpc$^{-1}$, $z^{*}=10$.

Curve 2 describes the dependence of $\Delta (m-M)$ on $z$,
assuming that evolutionary effects related to the mass of the
merging dwarfs and light absorption in galaxies play no role;
here, we adopted $\Omega_{\lambda}=0.7$, $\Omega_{M}=0.3$,
$\Omega_{tot}=1$, $H_{0}=70$ km s$^{-1}$ mpc$^{-1}$.

The line 3 corresponds to a model without evolutionary effects and
with no influence from dark energy ($\Omega_{\lambda}=0$,
$\Omega_{M}=0.2$, $\Omega_{tot}=0.2$, $H_{0}=70$ km s$^{-1}$
mpc$^{-1}$).

Curve 4 describes the evolution of light absorption by dust. This
relation was taken from \cite{kabanov2011} (Fig. 2 displaying the
dependence of the absorption $A_v$ on the redshift $z$, adopting
the curve corresponding to a disk galaxy with $M\sim R^2$).

Figure \ref{ris1} shows that dust absorption is able to attenuate
the brightness of SN Ia by as much as $\approx 1^{m}.3$. This
absorption is reached at redshifts $z\approx 3$. The evolution of
the mean mass of the merging CO WDs begins to be appreciable at
$z>2$ and leads to an increased brightness of SNe Ia in the past
compared to the current epoch. The mean apparent magnitude can
decrease by $\simeq 0^{m}.2$ at $z\gtrsim 8$, in full agreement
with the possibility that the mean mass of merging CO WDs (and,
according to our assumptions, the mean energy of SNe Ia) was
higher at the beginning of galaxy evolution, in the distant past,
compared to the current epoch.

To make it easier to compare our results to Figs. 4 and 5 from
\cite{riess1998a}, Fig. \ref{ris2} isolates the redshift range
between 0.01 and 1. The evolution of the mean mass of the merging
dwarfs essentially cannot influence the results in this redshift
range. Dust absorption attenuates the brightness of supernovae
much more strongly than if we allow for a possible contribution
from dark energy (Figs. 4 and 5 in \cite{riess1998a}), but recall
that \cite{kabanov2011} presents data for absorption by any dust,
including dust whose influence can be taken into account based on
reddening. In order to explain the accelerated fading of SNe Ia at
$z=0.5-1$ as due to evolution of gray-dust absorption in disk
galaxies, it is sufficient to suppose the amount of gray dust is
proportional to the total amount of gas in the host galaxies of
supernovae, and comprises several tens of percent of the total
dust absorption in these same galaxies.

\section{CONCLUSIONS}

We have considered two possible channels for the ``standard''
behavior of the SNe Ia brightness being violated. The first is the
time evolution of the mean mass of merging CO WDs, which are the
precursors of SNe Ia. The second is the time evolution of
absorption by gray dust in the SNe Ia host galaxies.

According to our computations, the mean energy of SNe Ia should
increase from $z>2$ and become significantly higher at $z \gtrsim
8$ due to the fact that, on average, mergers involved higher-mass
WDs during the early evolutionary stages of the Universe than at
the current epoch. Such distant supernovae have not yet been
discovered. Thus, the conclusion that the expansion of the
Universe is accelerating based on observations of SNe Ia at
redshifts to $z\simeq 1$ made in \cite{riess1998a,perlmutter1999a}
will not be affected by evolution of the mean mass of the merging
dwarfs. At the same time, it should always be borne in mind during
such estimation that our models, which are based on population
syntheses, assume that the objects whose evolution we are studying
(galaxies) were born at some time in the past (in the current
study, at $z^*=10$), then evolved. However, if a burst of star
formation occurred at some epoch after $z^*$, conclusions about
the increase of the mean SNe Ia energy should be related to the
time since that burst. Independent of the population's age, there
will exist some dispersion in supernova parameters, because the
combined mass of the merging dwarfs can vary within a factor of
approximately 1.5.

According to the model considered here, the evolution of radiation
absorption by galactic dust can give rise to a much stronger
attenuation of the brightness of distant supernovae than the
assumption that dark energy is present in the Universe. At the
same time, the absorption described by curve 4 in Figs. 1 and 2 is
the sum of the absorption that can, in principle, be taken into
account based on reddening and gray absorption, which cannot. By
varying the model using its intrinsic parameters and introducing
the required fraction of gray dust, we are able, in principle, to
achieve an essentially ideal agreement between curves 2 and 4 over
a wide redshift range (this reasoning is very similar to the
considerations presented in
\cite{aguirre1999a}-\cite{goobar2002}). However, this approach is
often criticized: fine tuning of the model parameters is believed
to be able to reproduce any results, and interpretation of the
phenomenon in this way seems unattractive \cite{riess2007}.

During the analysis of the SDSS-II supernova survey (and of the
light curves of these supernovae), SNe Ia were found to be
approximately $0^m.1$ brighter in quiet galaxies than in galaxies
with active star formation \cite{lampeitl2010}\footnote{Based on
analysis of supernovae with redshifts up to $z\simeq 0.2$ -- see
\cite{lampeitl2010}, Fig. 3.}. The characteristics of absorption
by dust are also different in galaxies with and without active
star formation \cite{lampeitl2010}. A dependence of the SNe Ia
luminosity on galaxy type was also noted in \cite{sullivan2010}.

We suggest the following interpretation of our results. Figure 2
shows that absorption by dust in disk galaxies grows at
essentially the same rate as the rate of dimming of supernovae due
to accelerated expansion of the Universe. Hence, we suggest that
the decrease in the observed brightness of supernova (on average,
and this is important) is due primarily to radiation absorption by
gray dust, whose influence cannot at this time be taken into
account using standard procedures for correcting for absorption.
However, further increase of the redshift results in increasing
absorption, causing still stronger brightness attenuation of
supernovae compared to a model without dust. At the same time,
some distant supernovae are found on the curve corresponding to
decelerating, rather than accelerating, expansion of the Universe
(see, for instance, Fig. 10 in \cite{kowalski2008}). This could be
considered unequivocal evidence against the idea that the observed
supernova fading can be explained by time-evolving absorption by
gray dust\footnote{Or, in principle, by any dust that was not duly
taken into account.} if all host galaxies of SNe Ia were disk
galaxies with active star formation. However, this is not the
case: for example, the parent galaxy of SN 1997ff (one of the most
distant supernovae, used as a standard) is elliptical
\cite{riess2001}.

As noted above, compared to disk galaxies, absorption in
elliptical galaxies is insignificant, and apparently does not
evolve over a considerable time interval, comparable to the Hubble
time. Note also that population syntheses of the evolution of
close binaries (e.g., Fig. 1 in \cite{jorgensen1997} or Fig. 2 in
\cite{mennekens2010}) show that the rate of SNe Ia outbursts in an
elliptical galaxy decreases by approximately a factor of two
between 1 and 10 billion years after the formation of the
elliptical galaxy. This means that the probability of observing a
SNe Ia in an elliptical galaxy, rather than a disk galaxy, was
much higher in the past (among supernovae with the highest $z$)
than at the current epoch (for low $z$). This can explain the
observed lack of ``accelerated'' supernovae for redshifts $z>1$.

Thus, when using SNe Ia as cosmological standard candles, it is
necessary to take into account not only model characteristics of
the supernovae, but also evolution of absorption by dust,
including gray dust, in the supernova host galaxies. It is
preferable to use supernovae in galaxies without star formation as
distance indicators\footnote{Note that some authors (e.g.,
\cite{vishwakarma2003}) believe that available observations,
including those for SNe Ia, are consistent with a wide range of
models of the Universe.}. Supernova outbursts in galaxies with
active current star formation can serve as probes for studies of
the evolution of absorption in such galaxies (including gray
absorption\footnote{Recall that gray absorption has been
discovered in the Milky Way, and can reach several tenths of a
magnitude \cite{gorbikov2010}.}).

Variations in the combined masses of binary degenerate dwarfs with
ages comparable to the age of the Universe do not significantly
change the brightness of SNe Ia explosions. It follows from the
discussions above that analysis of the evolution of SNe Ia
brightnesses aimed at improving cosmological models must be
carried out separately for elliptical and spiral galaxies. SNe Ia
in elliptical galaxies seem to provide a more reliable
cosmological tool than SNe Ia in spiral galaxies, since the latter
contain dust. Searches for and studies of very distant SNe Ia, at
redshifts $z>2$ are especially important.

\section{ACKNOWLEDGMENTS}

This work was supported by a grant from the President of the
Russian Federation for the State Support of Young Russian PhDs
(MK-142.2009.2), the Departmental Targeted Analytical Program "The
Development of the Scientific Potential of Higher Education"
(RNP-2.1.1/2906), the State Program of Support for Leading
Scientific Schools of the Russian Federation (NSh-7179.2010.2),
the Basic Research Program of the Department of Physical Studies
of the Russian Academy of Sciences "Active Processes and
Stochastic Structures in theUniverse," the Basic Research Program
of the Presidium of the Russian Academy of Sciences "Origin,
Structure, and Evolution of Objects in the Universe," the Russian
Foundation for BasicResearch (project 10-02-00231), and an
Integration project of the Siberian Division of the Russian
Academy of Sciences (No. 103).

\newpage

\section*{FIGURE CAPTIONS}

\begin{figure*}[h!]
\hspace{0cm} \epsfxsize=0.99\textwidth\centering
\epsfbox{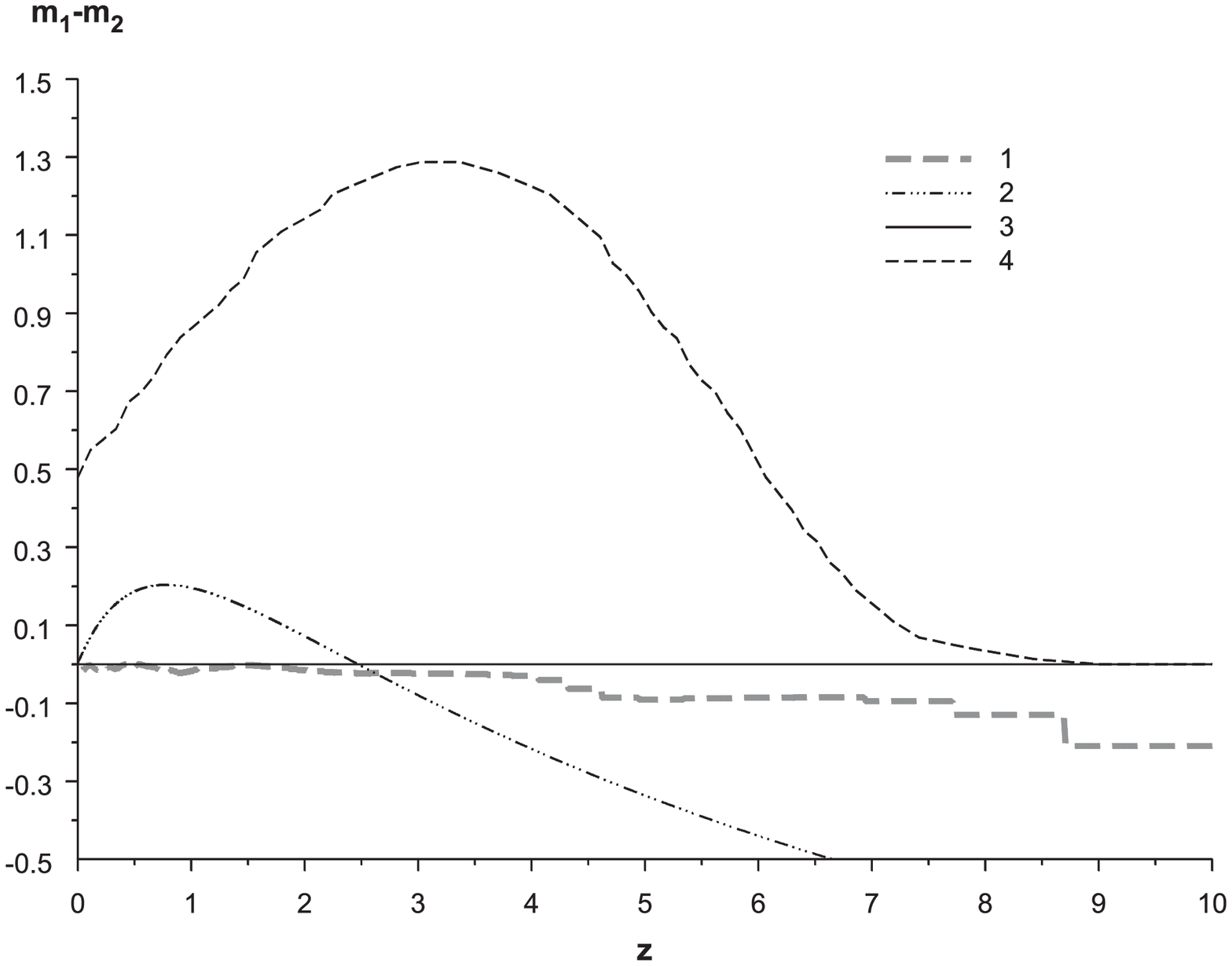} \vspace{0cm}\caption{} \label{ris1}
\end{figure*}

Fig. \ref{ris1}: Difference between the magnitude-redshift
dependences taking into account the effects studied here (time
evolution of the absorption and the mean mass of merging WDs) and
assuming standard SNe Ia that are not subject to variable
absorption in time, $\Delta (m-M)$. Curve 1 assumes that
absorption does not play any role and that the absolute magnitude
evolves in time from a maximum value at the beginning of the
evolution to a minimum, approximately constant, value after about
one billion years of stellar evolution in galaxies (for
$\Omega_{\lambda}=0$, $\Omega_{M}=0.2$, $\Omega_{tot}=0.2$,
$H_{0}=70$ km s$^{-1}$ Mpc$^{-1}$, $z^{*}=10$). Curve 2 assumes
that absorption does not play any role and the absolute magnitudes
experience no evolutionary effect ($\Omega_{\lambda}=0.7$,
$\Omega_{M}=0.3$, $\Omega_{tot}=1$, $H_{0}=70$ km s$^{-1}$
Mpc$^{-1}$). The line 3 corresponds to a model without
evolutionary effects ($\Omega_{\lambda}=0$, $\Omega_{M}=0.2$,
$\Omega_{tot}=0.2$, $H_{0}=70$ km s$^{-1}$ Mpc$^{-1}$). Curve 4
represents the evolution of light absorption by dust, taken from
\cite{kabanov2011} (Fig. 2 displaying the dependence of $A_v$ on
the redshift $z$; the curve corresponding to a disk galaxy with
$M\sim R^2$).

\bigskip

\begin{figure*}[h!]
\hspace{0cm} \epsfxsize=0.99\textwidth\centering
\epsfbox{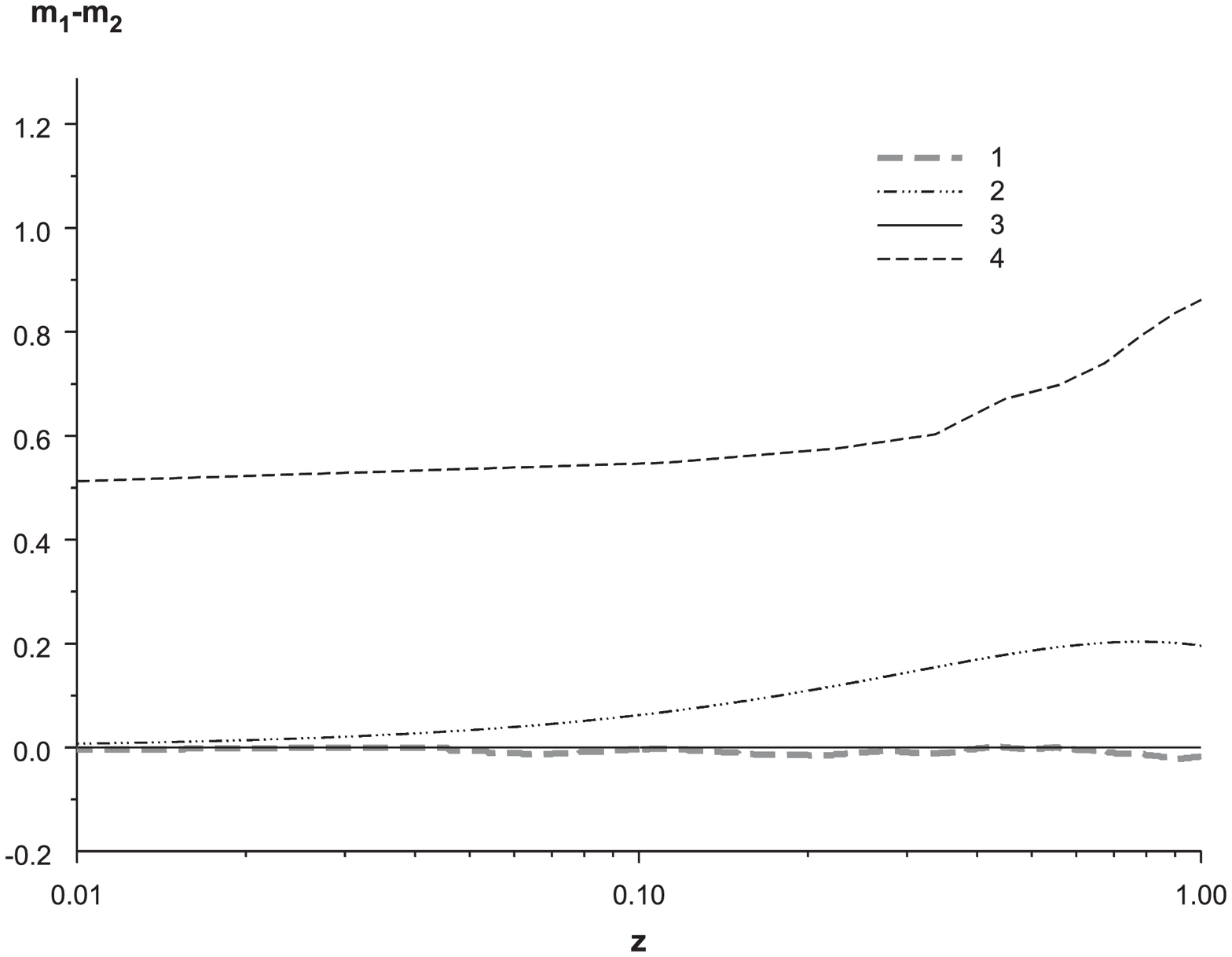} \vspace{0cm}\caption{} \label{ris2}
\end{figure*}

Fig. \ref{ris2}: Same as Fig. 1 for redshifts between 0.01 and 1.

\newpage

.

\newpage

\end{document}